\documentclass[final]{raa}           

\usepackage{graphicx,times}
\usepackage{natbib}
\usepackage{amssymb,amsmath}
\usepackage{caption}
\usepackage{subcaption}
\usepackage{scalerel}
\usepackage{graphicx}
\usepackage{lscape}
\usepackage{multirow}
\bibpunct{(}{)}{;}{a}{}{,}

\usepackage[pagebackref=true]{hyperref}

\usepackage{lineno}

\begin{document}

   \title{Studying the Equilibrium Points of the Modified Circular Restricted Three-Body Problem: the Case of Sun-Haumea System}

 \volnopage{ {\bf 20XX} Vol.\ {\bf X} No. {\bf XX}, 000--000}
   \setcounter{page}{1}

   \author{I. Nurul Huda\inst{1}, B. Dermawan\inst{2}, M. B. Saputra\inst{1}, R. Sadikin\inst{1}, T. Hidayat\inst{2}}

   \institute{ Research Center for Computing, National Research and Innovation Agency, Bogor, Indonesia\\
        \and
                Department of Astronomy and Bosscha Observatory, FMIPA, Institut Teknologi Bandung, Bandung, Indonesia\\
        {\it Email: ibnu.nurul.huda@brin.go.id} \\
\vs \no
   {\small Received 20XX Month Day; accepted 20XX Month Day}
}

\abstract{
We intend to study a modified version of the planar Circular Restricted Three-Body Problem (CRTBP) by incorporating several perturbing parameters. We consider the bigger primary as an oblate spheroid and emitting radiation while the small primary has an elongated body. We also consider the perturbation from a disk-like structure encompassing this three-body system. First, we develop a mathematical model of this modified CRTBP. We have found there exist five equilibrium points in this modified CRTBP model, where three of them are collinear and the other two are non-collinear. Second, we apply our modified CRTBP model to the Sun-Haumea system by considering several values of each perturbing parameter. Through our numerical investigation, we have discovered that the incorporation of perturbing parameters has resulted in a shift in the equilibrium point positions of the Sun-Haumea system compared to their positions in the classical CRTBP. The stability of equilibrium points is investigated. We have shown that the collinear equilibrium points are unstable and the stability of non-collinear equilibrium points depends on the mass parameter $\mu$ of the system. Unlike the classical case, non-collinear equilibrium points have both a maximum and minimum limit of $\mu$ for achieving stability. We remark that the stability range of $\mu$ in non-collinear equilibrium points depends on the perturbing parameters. In context of the Sun-Haumea system, we have found that the non-collinear equilibrium points are stable. 
}

   \authorrunning{Nurul Huda et al. }            
   \titlerunning{Studying the Equilibrium Points of the Modified CRTBP: the Case of Sun-Haumea System}  
   \maketitle

\section{Introduction}\label{sec1}

Celestial mechanics plays an important role in understanding the dynamics of Solar System Bodies \citep[see, e.g.,][]{murray1999solar,souchay2010dynamics,lei2021new,pan2022resonant}. One of the problems in celestial mechanics is the Circular Restricted Three-Body Problem (CRTBP). The study of CRTBP has aim to investigate the movement of an infinitesimal object under the gravitational influence of two primaries that have a circular orbit around their center of mass. CRTBP has several applications, such as for deep space exploration and satellite navigation.  The classical version of CRTBP assumes the primaries as a point mass and it only considers the gravitational interaction between them. There are five equilibrium points in the case of planar. Three of them are collinear ($L_1$, $L_2$, and $L_3$) and other two are non-collinear ($L_4$ and $L_5$) \citep{murray1999solar}. In order to make CRTBP model more realistic, the classical version has been modified by considering several additional parameters.

A stellar object, including the Sun, emits radiation. This radiation exerts pressure on objects in its path. There have been numerous studies that have considered radiation pressure force as another additional force in the restricted three-body problem \citep[see, e.g.,][]{haque1995non,ishwar2001secular,kushvah2007nonlinear,kushvah2008effect,das2009out,yousuf2019effects,patel2023analysis}. For instance, the first study on this topic has been done by \citet{radzievskii1950restricted}. \citet{chernikov1970photogravitational} extended the study by considering the relativistic Poynting-Robertson effect. \citet{simmons1985restricted} studied the effect of radiation pressure force in all ranges of value.  More recently, \citet{idrisi2017study} and \citet{idrisi2018non} considered the effect of planetary albedo on CRTBP as a consequence of solar radiation pressure force.

Since the stars and planets are not perfectly spherical, another aspect that has been considered in the CRTBP is the oblateness of the primaries. Early studies about the impact of an oblate primary on the dynamics of restricted three-body problem have been given by \citet{danby1965inclusion}, \citet{sharma1978case}, \citet{sharma1986finite}. More recently, the effect of oblateness on the dynamics of CRTBP has been studied in detail by several authors \citep[see, e.g.,][]{markellos1996non,douskos2006out,safiya2012oblateness,abouelmagd2013effect,zotos2015does,yousuf2022motion}. Moreover, some authors have considered the effect of both oblateness and radiation force in their calculation. For instance, \citet{singh1999stability} studied the linear stability of triangular equilibrium points when both primaries are oblate and emitting radiation. This study has been extended by \citet{singh2009combined} for the non-linear stability of $L_4$. \citet{abdulraheem2006combined} investigated the dynamics of CRTBP when both of primaries are oblate and emit radiation, together with the perturbation in the Coriolis and centrifugal force. Other authors such as \citet{huda2015locations}, \citet{dermawan2015triangular} and \citet{mia2023analysis}, have considered the effect of oblateness and radiation force in the Elliptic Restricted Three-Body Problem.

Our solar system contains several types of celestial bodies. Among them are elongated objects like a few asteroids, comets, and dwarf planets. These celestial bodies can be approximately described as finite straight segments. Previous studies of CRTBP have been enriched by assuming one or both primaries have an elongated body. At first, \citet{riaguas1999periodic} and \citet{riaguas2001non} analyzed the dynamics of a two-body problem by considering  one of the primaries as a finite straight segment. These works are extended by, e.g., \citet{jain2014stability}, \citet{kaur2020effect}, and \citet{kumar2019robe}, into the restricted three body-problem assuming both or one of the primaries have elongated shapes. In more recent studies, \citet{verma2023perturbed} examined the perturbed restricted three-body problem, where the smaller primary has an elongated shape and the larger primary is oblate and emits radiation. \citet{verma2023effect} considered the effect of finite straight segment and oblateness to study the dynamics of the restricted 2+2 body problem. 

Meanwhile, the effect of a disk-like structure as a perturbing force near a three-body system has been well studied by several authors \citep[see, e.g.,][]{chermnykh1987stability,jiang2004chaotic,kushvah2008linear,kushvah2012existence,kishor2013linear,mahato2022dynamics}. \citet{jiang2004chaotic} studied CRTBP by analyzing the influence of a disk-like structure near the three-body system. \citet{yousuf2019effects} analyzed the effect of disk-like structure, oblateness, and albedo on the CRTBP. \citet{mahato2022dynamics} extended the study of classical CRTBP by considering a disk-like structure and an elongated body. \citet{mahato2022effect} investigated the stability of equilibrium points within a framework of the perturbed restricted 2 + 2 bodies problem, taking into account the influence of a disk-like structure.

This study aims to obtain the collinear and non-collinear equilibrium points and investigate their stability under a framework of modified CRTBP incorporating the effect of radiation pressure, oblateness, finite straight segment, and disk-like structure. We intended to extend the work of \citet{yousuf2019effects} by assuming the small primary as a finite straight segment rather than oblate. It is also an extension of \citet{mahato2022dynamics} since we consider the effect of oblateness and radiation from the bigger primary.

Here we apply our modified CRTBP model to the Sun-Haumea system by assuming the Sun is a bigger primary with an oblate shape and emitting radiation and Haumea is a smaller primary which has an elongated body. We also consider the Kuiper belt as a disk-like structure surrounding the Sun-Haumea system. Haumea was chosen as our case study because of its unique characteristics, which have captured the attention of scientists since its discovery in 2003. The Haumea surface is dominantly covered by water ice \citep{barkume2006water, pinilla2009surface,noviello2022let}. There is also evidence that organic material exists on the Haumea's surface \citep{lacerda2008high,gourgeot2016near}. Recently, it has been discovered that the Haumea has a ring and two satellites named Namaka and Hi'iaka \citep{ortiz2017size}. Moreover, previous studies have proposed Haumea as a destination for space missions in the coming decades \citep[see, e.g.,][]{grundy2009exploration,sanchez2014optimal}. 

Besides the Sun-Haumea system, this modified CRTBP model can be applied to other cases. For instance, many planetary systems outside of our solar system have been discovered, and some systems have been found to have dust particle disks or asteroid belts, which are believed to be similar to the Kuiper belt or main belt in our solar system \citep[see, e.g.,][]{greaves1998dust,matra2019kuiper}. Meanwhile, previous studies have explained the presence of extrasolar asteroids or dwarf planets near the host star \citep[see, e.g.,][]{jura2003tidally,dufour2010discovery}. Moreover, some space explorations have been devoted to exploring small solar system bodies near the main belt or Kuiper belt region. It is known that several solar system bodies have an irregular shape. Therefore, it is reasonable to study the combined effects of perturbations from a disk, an elongated body, and an oblate radiating body on the motion of an infinitesimal mass in the CRTBP.  

The structure of this paper is as follows. In the next section, we present a mathematical formulation of the dynamical model. The position and the stability of equilibrium points are given in Section \ref{sec3}. Section \ref{sec4} gives the implementation of the dynamical model in the Sun-Haumea system. Finally, the conclusion is given in Section \ref{sec5}. Here, MATLAB's Symbolic Toolbox is used to conduct certain algebraic calculations and find numerical solutions.

\section{Mathematical formulation of the dynamical system}\label{sec2}

In this work, we consider a system where an infinitesimal mass moves under the influence of a bigger primary with mass $m_1$ and a small primary with mass $m_2$. The primaries of this system have a circular orbit around their center of mass. We treat the bigger primary as a source of radiation with an oblate spheroid shape, while the small primary has an elongated shape. The unit of time is normalized to make the Gaussian constant of gravitation equal to one. The mass parameter is represented by $\mu = m_2/(m_1 + m_2)$ where $m_1 = 1-\mu$ and $m_2 = \mu$. In the case of a restricted three-body problem, it is more convenient to introduce the system in the rotational coordinate $Oxy$. The primaries are located in the $x$-axis with the distance between primaries chosen as the unit of length. The coordinates of the bigger primary, small primary, and the third body are ($\mu, 0$), ($\mu-1, 0$), and ($x, y$), respectively. The oblateness factor of the bigger primary can be represented by $A = (AE^2 - AP^2)/5R^2$ where $A \ll 1$, $AE$ and $AP$ represent the equatorial and polar radii, respectively, and $R$ is the effective radius when assuming the primary as a spherical object. Meanwhile, the radiation force $F_p$ acts opposite to gravitational force and diminishes with respect to the distance. The total force acting on the bigger primary can be written as $F_g - F_p = qF_g$, hence $q = 1 - (F_p/F_g)$. Here $q$ is called the mass reduction factor where $0 < 1-q \ll 1$. The small primary is assumed as a finite straight segment with a length $2l$. The effect of a disk-like structure surrounding the system is also considered in this study. Following \citet{miyamoto1975three}, the planar version of unit less potential disk-like structure is given by $V(x,y) = M_b/\sqrt{r^2 + T^2}$, where $M_b$ is the total mass of disk-like structure, $r^2 = x^2 + y^2$ is the radial distance of the infinitesimal mass, $T = a + b$ is the total of flatness and core parameters. Let the distance of primaries to the center of mass are $s_1$ and $s_2$. Considering the previous works such as \citet{kushvah2008linear},\citet{yousuf2019effects}, and \citet{mahato2022dynamics}, the motion of the primaries is given by 
\begin{align} \label{eq:meanmotion0}
\begin{split}
m_1s_1n^2 = \frac{Gm_1m_2}{R^2-l^2}\left(1+\frac{3A}{2R^2}\right)+\frac{GM_bm_1r_c}{(r^2_c+T^2)^{3/2}}, \\
m_2s_2n^2 = \frac{Gm_1m_2}{R^2-l^2}\left(1+\frac{3A}{2R^2}\right)+\frac{GM_bm_2r_c}{(r^2_c+T^2)^{3/2}},
\end{split}
\end{align}
where $R = s_1 + s_2$ is the distance between primaries. $r_c^2 = 1 - \mu + \mu^2$ means a dimensionless quantity of the reference radius of the disk-like structure \citep{singh2014effects}. Assuming $R = 1$, $G = 1$, and $m_1 + m_2 = 1$, the mean motion $n$ of the system can be calculated by adding both equations in Eq. \ref{eq:meanmotion0}, approximating the expression $1/(1-l^2)$ in series as $1+l^2$, and neglecting the term of $Al^2$. Hence we have:
\begin{equation} \label{eq:meanmotion}
n^2 = 1 + l^2 + \frac{3}{2} A + \frac{2 M_b r_c}{(r^2_c + T^2)^{3/2}}
\end{equation}
Equations of motion of the third object in CRTBP are given as follows
\begin{align} \label{eq:sirkular-gerak-eta}
\begin{split}
\ddot{x} - 2n\dot{y} = \frac{\partial \Omega}{\partial x}, \\
\ddot{y} + 2n\dot{x} = \frac{\partial \Omega}{\partial y}, 
\end{split}
\end{align}
where $U$ is a pseudo-potential function:
\begin{equation} \label{eq:sirkular-potensial}
\begin{aligned}
\Omega = {} & \frac{n^2}{2}\left(x^2 + y^2\right) + \frac{q(1 - \mu)}{r_1} + \frac{(1 - \mu)A q}{r_1^3} \\
       & + \frac{\mu}{2l}\log\left(\frac{r_{21}+r_{22}+2l}{r_{21}+r_{22}-2l}\right) + \frac{M_b}{\sqrt{r^2 + T^2}}.
\end{aligned}
\end{equation}
Here $r^2_{21} = ( x - \mu + 1 - l)^2 + y^2$ and $r^2_{22} = ( x - \mu + 1 + l)^2 + y^2$ are the distance of third body to the small primary and $r^2_1 = (x-\mu)^2 + y^2$ is the distance between third body and the bigger primary. It should be noted that the equation of motion differs from the equation of motion in \citet{yousuf2019effects} since, in our case, we assume the small body by a finite straight segment.

\section{Equilibrium points} \label{sec3}

\subsection{Position of equilibrium points}

The conditions of equilibrium points are $\dot{x}=\dot{y}=\ddot{x}=\ddot{y} = 0$. Hence we can deduce that $\Omega_x = \Omega_y = 0$, i.e.,

\begin{equation} \label{eq:u_x}
\begin{aligned}
n^2x - {}& \frac{q(1-\mu)(x-\mu)}{r^3_1} - \frac{3(1 - \mu)(x - \mu)A q}{2r_1^5} \\
& - \frac{2\mu}{(r_{21}+r_{22})^2-4l^2}\left(\frac{x-\mu+1-l}{r_{21}}+\frac{x-\mu+1+l}{r_{22}}\right) \\
& - \frac{M_bx}{(r^2 + T^2)^{3/2}} = 0, \\
\end{aligned}
\end{equation}

\begin{equation} \label{eq:u_y}
\begin{aligned}
n^2y - {}& \frac{q(1-\mu)y}{r^3_1} - \frac{3(1 - \mu)y A q}{2r_1^5} \\
& - \frac{2\mu}{(r_{21}+r_{22})^2-4l^2}\left(\frac{y}{r_{21}}+\frac{y}{r_{22}}\right) - \frac{M_by}{(r^2 + T^2)^{3/2}} = 0. \\
\end{aligned}
\end{equation}
In the following, we solve Eq. \ref{eq:u_x} and Eq. \ref{eq:u_y} to find the position of equilibrium points.

The collinear points are located in a line with the primaries, thus we have $y = 0$. Eq. \ref{eq:u_x} becomes
\begin{align} \label{eq:u_x0}
\normalsize
\begin{split}
\Omega_x(x,0) = {}& n^2x - \frac{q(1-\mu)(x-\mu)}{\lvert x-\mu \rvert^3} - \frac{3(1 - \mu)(x - \mu)A q}{2\lvert x-\mu\rvert^5} \\
& - \frac{2\mu}{(\lvert x-\mu+1+l\rvert + \lvert x-\mu+1-l \rvert)^2-4l^2} \\
& \times \left(\frac{x-\mu+1-l}{\lvert x-\mu+1-l\rvert} + \frac{x-\mu+1+l}{\lvert x-\mu+1+l\rvert}\right) \\
& - \frac{M_b}{x^2}\left(1-\frac{3T^2}{2x^2}\right) = 0.
\end{split}
\end{align}
In order to find the solution, we divide the region into three parts, i.e. ($-\infty$, $\mu-1-l$), ($\mu-1-l$, $\mu$), and ($\mu$, $\infty$). Here $L_1$, $L_2$, and $L_3$ are the solution located in ($-\infty$, $\mu-1-l$), ($\mu-1-l$, $\mu$), and ($\mu$, $\infty$), respectively. Hence we have
\begin{align} \label{eq:u_x0_cond}
    \Omega_x(x,0) = 
\begin{cases}
    \left(n^2x - \frac{M_b}{x^2}\left(1-\frac{3T^2}{2x^2}\right) \right) (x-\mu)^4((2x-2\mu+2)^2-4l^2) \\
    \qquad + q(1-\mu)(x-\mu)^2((2x-2\mu+2)^2-4l^2)  \\
    \qquad + \frac{3}{2}qA(1-\mu)((2x-2\mu+2)^2-4l^2) + 4\mu(x-\mu)^4, \\
    \vspace{5mm}
    \qquad \text{if } -\infty < x <  \mu-1-l\\ 
    \left(n^2x - \frac{M_b}{x^2}\left(1-\frac{3T^2}{2x^2}\right) \right) (x-\mu)^4 ((2x-2\mu+2)^2-4l^2) \\
    \qquad + q(1-\mu)(x-\mu)^2((2x-2\mu+2)^2-4l^2) \\
    \qquad + \frac{3}{2}qA(1-\mu)((2x-2\mu+2)^2-4l^2) - 4\mu(x-\mu)^4, \\
    \vspace{5mm}
    \qquad \text{if } \mu-1+l < x <  \mu\\
    \left(n^2x - \frac{M_b}{x^2}\left(1-\frac{3T^2}{2x^2}\right) \right) (x-\mu)^4 ((2x-2\mu+2)^2-4l^2) \\
    \qquad - q(1-\mu)(x-\mu)^2((2x-2\mu+2)^2-4l^2) \\
    \qquad - \frac{3}{2}qA(1-\mu)((2x-2\mu+2)^2-4l^2) - 4\mu(x-\mu)^4. \\
    \qquad \text{if } \mu < x <  \infty
\end{cases}
\end{align}
These three equations have been solved numerically to find each collinear equilibrium point. Only the real solution is considered for the position of equilibrium points. 

Meanwhile, there are two non-collinear equilibrium points, i.e., $L_4$ and $L_5$. The additional condition of these equilibrium points is $y \neq 0$. Eq. \ref{eq:u_x} and \ref{eq:u_y} can be rewritten in the form
\begin{align} \label{eq:u_x_1}
\footnotesize
\begin{split}
x \left( n^2 - \frac{q(1-\mu)}{r^3_1} - \frac{3(1 - \mu)A q}{2r_1^5} - \frac{2\mu}{(r_{21}+r_{22})^2-4l^2}\left(\frac{1}{r_{21}}+\frac{1}{r_{22}}\right) \right. {}& \\
\left. - \frac{M_bx}{(r^2 + T^2)^{3/2}} \right) \frac{q\mu(1-\mu)}{r^3_1} + \frac{3\mu(1 - \mu)A q}{2r_1^5} {}& \\
- \frac{2\mu}{(r_{21}+r_{22})^2-4l^2}\left(\frac{-\mu+1-l}{r_{21}}+\frac{-\mu+1+l}{r_{22}}\right) {}& = 0.
\end{split}
\end{align}
\begin{align} \label{eq:u_y_1}
\footnotesize
\begin{split}
y\left(n^2 - \frac{q(1-\mu)}{r^3_1} - \frac{3(1 - \mu)A q}{2r_1^5} - \frac{2\mu}{(r_{21}+r_{22})^2-4l^2}\left(\frac{1}{r_{21}}+\frac{1}{r_{22}}\right) \right. {}& \\
\left. - \frac{M_by}{(r^2 + T^2)^{3/2}}\right) = 0. 
\end{split}
\end{align}
Hence from Eq. \ref{eq:u_y_1} we have
\begin{align} \label{eq:u_y_2}
\footnotesize
\begin{split}
n^2 - \frac{q(1-\mu)}{r^3_1} - \frac{3(1 - \mu)A q}{2r_1^5} - \frac{2\mu}{(r_{21}+r_{22})^2-4l^2}\left(\frac{1}{r_{21}}+\frac{1}{r_{22}}\right) - \frac{M_by}{(r^2 + T^2)^{3/2}} = 0. 
\end{split}
\end{align}
Substituting Eq. \ref{eq:u_y_2} into Eq. \ref{eq:u_x_1} gives
\begin{align} \label{eq:u_x_2}
\footnotesize
\begin{split}
\frac{q(1-\mu)}{r^3_1} + \frac{3(1 - \mu)A q}{2r_1^5} - \frac{2}{(r_{21}+r_{22})^2-4l^2}\left(\frac{-\mu+1-l}{r_{21}}+\frac{-\mu+1+l}{r_{22}}\right) = 0.
\end{split}
\end{align}

In the classical case, the position of these equilibrium points is located in $r_1 = 1$ and $r_2 = 1$. Since some perturbations exist, we assume that $r_1$ and $r_2$ are perturbed by $\epsilon_1$ and $\epsilon_2$. Hence, in our case, we have \citep{mahato2022dynamics}
\begin{align} \label{eq:r_perturbed}
r_1 = 1 + \epsilon_1; \qquad r_{21} = 1 + \epsilon_2 - l/2; \qquad r_{22} = 1 + \epsilon_2 + l/2. 
\end{align}
The calculation of $\epsilon_1$ and $\epsilon_2$ are done by substituting Eq. \ref{eq:r_perturbed} to Eq. \ref{eq:u_x_2} and Eq. \ref{eq:u_y_2} and solving these equations. By approximating with series and neglecting higher order terms of $\epsilon_1$, $\epsilon_2$, $l^2$, and $A$, we have:
\begin{align} \label{eq:epsilon}
\begin{split}
\epsilon_1 &= \frac{\frac{4\,\gamma }{3}-\frac{4\,q }{3}-2\,A \,q +\frac{\gamma \,\mu }{3}+\frac{4\,\mu \,q }{3}+2\,A \,\mu \,q }{4\,q \,{\left(\mu -1\right)}} \\
& +\frac{5\,\gamma \,\mu -l^2 \,{\left(\frac{2\,\mu }{3} +\frac{11\,\gamma \,\mu }{4}\right)}}{q \,{\left(13\,l^2 -12\right)}\,{\left(\mu -1\right)}}, \\
\epsilon_2 &= \frac{16\,\gamma -40\,l^2 \,\gamma +28\,l^2 -16}{52\,l^2 -48}
\end{split}
\end{align}
where $\gamma = 1 + l^2 + 3A/2 +  M_b(2r_c-1)/(r_c^2+T^2)^{3/2}$. The position of non-collinear equilibrium points ($x_o,y_o$) is given by
\begin{align} \label{eq:noncol_eq}
\begin{split}
    x_o &= \mu - \frac{1}{2}+\left(\epsilon_2-\epsilon_1\right), \\
    y_o &= \pm \sqrt{\frac{3}{4}+\epsilon_1 + \epsilon_2}
\end{split}
\end{align}

Putting value of $\epsilon_{1,2}$ into Eq. \ref{eq:noncol_eq}, we get:
\begin{align} \label{eq:noncol_eq_fix}
\footnotesize
\begin{split}
    x_o &= \mu -\frac{1}{2} +\frac{16\,\gamma -40\,L^2 \,\gamma +28\,L^2 -16}{52\,L^2 -48} \\
    & -\frac{\frac{4\,\gamma }{3}-\frac{4\,q_1 }{3}-2\,A_1 \,q_1 +\frac{\gamma \,\mu }{3}+\frac{4\,\mu \,q_1 }{3}+2\,A_1 \,\mu \,q_1 }{4\,q_1 \,{\left(\mu -1\right)}}-\frac{5\,\gamma \,\mu -L^2 \,{\left(\frac{2\,\mu }{3}+\frac{11\,\gamma \,\mu }{4}\right)}}{q_1 \,{\left(13\,L^2 -12\right)}\,{\left(\mu -1\right)}}, \\
    y_o &= \pm \left(\frac{3}{4}+\frac{16\,\gamma -40\,L^2 \,\gamma +28\,L^2 -16}{52\,L^2 -48} \right. \\
    & \left. +\frac{\frac{4\,\gamma }{3}-\frac{4\,q_1 }{3}-2\,A_1 \,q_1 +\frac{\gamma \,\mu }{3}+\frac{4\,\mu \,q_1 }{3}+2\,A_1 \,\mu \,q_1 }{4\,q_1 \,{\left(\mu -1\right)}}+\frac{5\,\gamma \,\mu -L^2 \,{\left(\frac{2\,\mu }{3}+\frac{11\,\gamma \,\mu }{4}\right)}}{q_1 \,{\left(13\,L^2 -12\right)}\,{\left(\mu -1\right)}}\right)^{1/2}
\end{split}
\end{align}
If the perturbation parameters are not considered, Eq. \ref{eq:noncol_eq_fix} is similar to the classical version where $x_o = \mu - \frac{1}{2}$ and $y_o = \pm \sqrt{\frac{3}{4}}$.

\subsection{Linear Stability}

Let us assume a small displacement in an equilibrium point by defining
\begin{equation} \label{eq:displacement}
    u = x - x_o; \qquad v = y - y_o,
\end{equation}
where "$o$" corresponds to the equilibrium points. The equation of motion from this small displacement is given as follows:
\begin{align} \label{eq:motion_displacement}
\begin{split}
\ddot{u} - 2n\dot{v} = u\Omega^o_{xx} + v\Omega^o_{xy}, \\
\ddot{v} + 2n\dot{u} = u\Omega^o_{xy} + v\Omega^o_{yy}, \\
\end{split}
\end{align}
where
\begin{equation} \label{eq:u_xx}
\begin{aligned}
\scaleto{\Omega^o_{xx} =}{6pt}{}& \scaleto{n^2 + \frac{3q(1-\mu)(x-\mu)^2}{r^5_1} - \frac{q(1-\mu)}{r^3_1} + \frac{15A q(1 - \mu)(x - \mu)^2}{2r_1^7} - \frac{3(1 - \mu)A q}{2r_1^5} + \frac{3M_bx^2}{(r^2 + T^2)^{5/2}}}{16pt} \\
& \scaleto{- \frac{M_b}{(r^2 + T^2)^{3/2}} + \frac{2\mu}{(r_{21}+r_{22})^2-4l^2}\left(\frac{1}{r_{21}+r_{22}-2l}+\frac{1}{r_{21}+r_{22}+2l}\right)\left(\frac{x-\mu+1-l}{r_{21}}+\frac{x-\mu+1+l}{r_{22}}\right)^2 }{16pt} \\
& \scaleto{- \frac{2\mu}{(r_{21}+r_{22})^2-4l^2}\left[\frac{1}{r_{21}}+\frac{1}{r_{22}}-\left(\frac{(x-\mu+1-l)^2}{r^3_{21}}+\frac{(x-\mu+1+l)^2}{r^3_{22}}\right)\right]}{16pt},
\end{aligned}
\end{equation}
\begin{equation} \label{eq:u_yy}
\begin{aligned}
\scaleto{\Omega^o_{yy} =}{7pt}{}& \scaleto{n^2 + \frac{3q(1-\mu)y^2}{r^5_1} - \frac{q(1-\mu)}{r^3_1} + \frac{15A q(1 - \mu)y^2}{2r_1^7} - \frac{3(1 - \mu)A q}{2r_1^5} + \frac{3M_by^2}{(r^2 + T^2)^{5/2}}- \frac{M_b}{(r^2 + T^2)^{3/2}}}{16pt} \\
& \scaleto{ + \frac{2\mu}{(r_{21}+r_{22})^2-4l^2}\left(\frac{1}{r_{21}+r_{22}-2l}+\frac{1}{r_{21}+r_{22}+2l}\right)\left(\frac{y}{r_{21}}+\frac{y}{r_{22}}\right)^2 }{16pt} \\
& \scaleto{- \frac{2\mu}{(r_{21}+r_{22})^2-4l^2}\left[\frac{1}{r_{21}}+\frac{1}{r_{22}}-\left(\frac{y^2}{r^3_{21}}+\frac{y^2}{r^3_{22}}\right)\right]}{16pt},
\end{aligned}
\end{equation}
\begin{equation} \label{eq:u_xy}
\begin{aligned}
\scaleto{\Omega^o_{xy} =}{6pt}{}& \scaleto{\frac{3q(1-\mu)(x-\mu)y}{r^5_1} + \frac{15A q(1 - \mu)(x - \mu)y}{2r_1^7} + \frac{3M_bxy}{(r^2 + T^2)^{5/2}}}{16pt} \\
& \scaleto{ + \frac{2\mu y}{(r_{21}+r_{22})^2-4l^2}\left(\frac{1}{r_{21}+r_{22}-2l}+\frac{1}{r_{21}+r_{22}+2l}\right)\left(\frac{1}{r_{21}}+\frac{1}{r_{22}}\right)\left(\frac{x-\mu+1-l}{r_{21}}+\frac{x-\mu+1+l}{r_{22}}\right) }{16pt} \\
& \scaleto{+ \frac{2\mu y}{(r_{21}+r_{22})^2-4l^2}\left(\frac{x-\mu+1-l}{r^3_{21}}+\frac{x-\mu+1+l}{r^3_{22}}\right)}{16pt}.
\end{aligned}
\end{equation}
Here $\Omega^o$ means the pseudo-potential is evaluated in equilibrium points. Hence it is constant. Eq. \ref{eq:motion_displacement} has general solutions
\begin{align} \label{eq:solution}
\begin{split}
u = \sum_{i = 1}^{4} \alpha_i e^{\lambda_{i}t}, \\
v = \sum_{i = 1}^{4} \beta_i e^{\lambda_{i}t}.
\end{split}
\end{align}
where $\alpha_i$ and $\beta_i$ are constants while $\lambda_{i}$ is the root of the characteristic equation. Substituting Eq. \ref{eq:solution} to Eq. \ref{eq:motion_displacement} produces
\begin{equation}\label{eq:matrices_eq}
   \begin{bmatrix}
     \lambda^2-\Omega^o_{xx} & -2n\lambda-\Omega^o_{xy}\\
     2n\lambda-\Omega^o_{xy} & \lambda^2-\Omega^o_{yy}
   \end{bmatrix}
   \begin{bmatrix}\alpha \\\beta \end{bmatrix}=
   \begin{bmatrix}0 \\0 \end{bmatrix}
\end{equation}
The first term of the left-hand side has to be a singular matrix. Hence the determinant of this matrix has to be zero:
\begin{equation} \label{eq:characteristic}
{\lambda}^4 + (4n^2 - \Omega^o_{xx} - \Omega^o_{yy})\lambda^2 + \Omega^o_{xx}\Omega^o_{yy} - (\Omega^o_{xy})^2 = 0.
\end{equation}
This equation is called the characteristic equation. It is a quadratic equation in $\lambda^2$. The solution of this quadratic equation is
$\lambda_i = \pm \sqrt{(-b \pm \sqrt{b^2 - 4c})/2}$, where $b = 4n^2 - \Omega^o_{xx} - \Omega^o_{yy}$ and $c = \Omega^o_{xx}\Omega^o_{yy} - (\Omega^o_{xy})^2$. If all obtained $\lambda_i$ are purely imaginary, then it gives the motion of stable periodic behaviour near the vicinity of equilibrium points. However, if there is at least one $\lambda_i$ which has a form of real or complex, then the third body is unstable since $u$ and $v$ will exponentially increase with respect to time. We can investigate the stability behaviour of the system by looking at the sign of $b$ and $c$. The system is stable if $b > 0$, $b^2-4c>0$, and $b>\sqrt{b^2-4c}$ since it produces all pure imaginary $\lambda_i$. 

\section{The Case of  Sun-Haumea System}\label{sec4}

In this work, we model the Sun-Haumea system through the framework of the restricted three-body problem with Sun as a bigger primary and Haumea as a small primary. Here we also consider the Kuiper belt in this system. We assume Haumea has a circular orbit and orbits in the same plane as the Kuiper belt. The mass of the small primary is a combination of Haumea's mass and the mass of Haumea's satellites: Namaka and Hi'iaka. Sun has a mass around $1.989 \times 10^{30}$ kg. Haumea has a length of $\sim 2300$ km for its largest axis and a mass of $4 \times 10^{21}$ kg \citep{ragozzine2009orbits}. Meanwhile, Namaka and Hi'iaka have a mass of $1.79 \times 10^{18}$ kg and $17.9 \times 10^{18}$ kg, respectively \citep{ortiz2017size}. Hence we have $\mu = 2 \times 10^{-9}$ and $l = 3.5 \times 10^{-7}$. Following \citet{yousuf2019effects}, here we assume that the Sun has $A = 2.6 \times 10^{-11}$ while the Kuiper belt has $T = 0.11$ and $M_b = 3 \times 10^{-7}$. According to \citet{sharma1987linear}, the photogravitational parameter $q$ can be expressed in the CGS unit system as $q = 1 - (5.6 \times 10^{-5}/a\rho)$ where $a$ and $\rho$ are the radius and density of a moving body, respectively. Assumed a spacecraft has $a = 700$ cm and $\rho = 0.05$ gr/cm$^3$, hence $1-q = 1.6 \times 10^{-6}$.

\begin{table}
\caption{The abscissa Position of collinear equilibrium points ($L_1$, $L_2$, and $L_3$) in Sun-Haumea system with $\mu = 2 \times 10^{-9}$ and $T = 0.11$. }
\begin{center}
{\renewcommand{\arraystretch}{1.1}
\setlength{\tabcolsep}{1.2pt}
\begin{tabular}{|c|c|c|c|c|c|c|}
\hline
$1-q$   & $A$   & $l$   & $M_b$     & $L_{1}$ & $L_{2}$ & $L_{3}$  \\ 
\hline
1     & 0     & 0     & 0      & $-1.000873832771965$ & $-0.999126671989864$ & $1.000000000833333$  \\
\hline
$1.6 \times 10^{-6}$     & $2.6 \times 10^{-11}$  & $3.5 \times 10^{-7}$ & $3 \times 10^{-7}$  & $-1.00087355691303$ & $-0.999126395886954$ & $0.999999369260791$  \\
$1.6 \times 10^{-4}$     & $2.6 \times 10^{-11}$  & $3.5 \times 10^{-7}$ & $3 \times 10^{-7}$  & $-1.00085632697634$ & $-0.999108413962744$ & $0.999946566436973$  \\
$1.6 \times 10^{-9}$     & $2.6 \times 10^{-11}$  & $3.5 \times 10^{-7}$ & $3 \times 10^{-7}$  & $-1.00087373433354$ & $-0.999126573666542$ & $0.999999902060868$  \\
\hline
$1.6 \times 10^{-6}$     & $2.6 \times 10^{-11}$  & $3.5 \times 10^{-7}$ & $3 \times 10^{-7}$  & $-1.00087355691303$ & $-0.999126395886954$ & $0.999999369260791$  \\
$1.6 \times 10^{-6}$     & $2.6 \times 10^{-9}$   & $3.5 \times 10^{-7}$ & $3 \times 10^{-7}$  & $-1.00087355691116$ & $-0.999126395888830$ & $0.999999369260793$  \\
$1.6 \times 10^{-6}$     & $2.6 \times 10^{-13}$  & $3.5 \times 10^{-7}$ & $3 \times 10^{-7}$  & $-1.00087355691305$ & $-0.999126395886936$ & $0.999999369260791$  \\
\hline
$1.6 \times 10^{-6}$     & $2.6 \times 10^{-11}$  & $3.5 \times 10^{-7}$ & $3 \times 10^{-7}$  & $-1.00087355691303$ & $-0.999126395886954$ & $0.999999369260791$  \\
$1.6 \times 10^{-6}$     & $2.6 \times 10^{-11}$  & $3.5 \times 10^{-5}$ & $3 \times 10^{-7}$  & $-1.00087402445000$ & $-0.999125928668234$ & $0.999999368852499$  \\
$1.6 \times 10^{-6}$     & $2.6 \times 10^{-11}$  & $3.5 \times 10^{-9}$ & $3 \times 10^{-7}$  & $-1.00087355686628$ & $-0.999126395933676$ & $0.999999369260832$  \\
\hline
$1.6 \times 10^{-6}$     & $2.6 \times 10^{-11}$  & $3.5 \times 10^{-7}$ & $3 \times 10^{-7}$  & $-1.00087355691303$ & $-0.999126395886954$ & $0.999999369260791$  \\
$1.6 \times 10^{-6}$     & $2.6 \times 10^{-11}$  & $3.5 \times 10^{-7}$ & $3 \times 10^{-5}$  & $-1.00086393713086$ & $-0.999116570293592$ & $0.999989644085257$  \\
$1.6 \times 10^{-6}$     & $2.6 \times 10^{-11}$  & $3.5 \times 10^{-7}$ & $3 \times 10^{-9}$  & $-1.00087365419779$ & $-0.999126493044322$ & $0.999999466517286$  \\
\hline
\end{tabular}}
\label{tab:eqtc}
\end{center}
\end{table}

We calculated the position of the collinear equilibrium points of Sun-Haumea. By substituting the property of the system into Eq. \ref{eq:u_x0_cond} and solving it numerically, we found $L_1$, $L_2$, and $L_3$.  Table \ref{tab:eqtc} shows the position of collinear equilibrium points. Here we vary the value of each perturbation parameter to examine the impact on the equilibrium point position. In the case of $L_1$, the position is getting closer to the primaries if $A$ and $1-q$ increase. Decreasing $A$ and increasing $1-q$ makes $L_2$ closer to the bigger primary. The position of $L_3$ is nearer with respect to primaries if the bigger primary emits stronger radiation pressure. According to Table \ref{tab:eqtc}, the position of collinear equilibrium points depends on the value of $M_b$ and $l$. Increasing $M_b$ and decreasing $l$ makes the location of $L_1$ nearer the smaller primary. The increment of $M_b$ and $l$ affect the position of $L_2$ to become closer to the bigger primary. $L_3$ is getting closer to the primaries if we increase the value of $M_b$.

\begin{table}
\caption{Position of non-collinear equilibrium points ($L_4$ and $L_5$) in Sun-Haumea system with $\mu = 2 \times 10^{-9}$ and $T = 0.11$. }
\begin{center}
{\renewcommand{\arraystretch}{1}
\begin{tabular}{|c|c|c|c|c|c|}
\hline
\multirow{2}{*}{$1-q$}   & \multirow{2}{*}{$A$}   & \multirow{2}{*}{$l$}   & \multirow{2}{*}{$M_b$}     & \multicolumn{2}{|c|}{$L_{4,5}$} \\ \cline{5-6}
        &       &       &           & $x$      & $y$ \\
\hline
1     & 0     & 0     & 0      & $-0.499999998000000$ & $\pm 	0.866025403784439$  \\
\hline
$1.6 \times 10^{-6}$     & $2.6 \times 10^{-11}$  & $3.5 \times 10^{-7}$ & $3 \times 10^{-7}$  & $-0.499999464678626$ & $\pm 0.866024982450545$  \\
$1.6 \times 10^{-4}$     & $2.6 \times 10^{-11}$  & $3.5 \times 10^{-7}$ & $3 \times 10^{-7}$  & $-0.499946656129219$ & $\pm 0.865994492868782$  \\
$1.6 \times 10^{-9}$     & $2.6 \times 10^{-11}$  & $3.5 \times 10^{-7}$ & $3 \times 10^{-7}$  & $-0.499999997479636$ & $\pm 0.866025290063446$  \\
\hline
$1.6 \times 10^{-6}$     & $2.6 \times 10^{-11}$  & $3.5 \times 10^{-7}$ & $3 \times 10^{-7}$  & $-0.499999464678626$ & $\pm 0.866024982450545$  \\
$1.6 \times 10^{-6}$     & $2.6 \times 10^{-9}$   & $3.5 \times 10^{-7}$ & $3 \times 10^{-7}$  & $-0.499999465965624$ & $\pm 0.866024981707493$  \\
$1.6 \times 10^{-6}$     & $2.6 \times 10^{-13}$  & $3.5 \times 10^{-7}$ & $3 \times 10^{-7}$  & $-0.499999464665756$ & $\pm 0.866024982457975$  \\
\hline
$1.6 \times 10^{-6}$     & $2.6 \times 10^{-11}$  & $3.5 \times 10^{-7}$ & $3 \times 10^{-7}$  & $-0.499999464678626$ & $\pm 0.866024982450545$  \\
$1.6 \times 10^{-6}$     & $2.6 \times 10^{-11}$  & $3.5 \times 10^{-5}$ & $3 \times 10^{-7}$  & $-0.499999464372405$ & $\pm 0.866024982155884$  \\
$1.6 \times 10^{-6}$     & $2.6 \times 10^{-11}$  & $3.5 \times 10^{-9}$ & $3 \times 10^{-7}$  & $-0.499999464678656$ & $\pm 0.866024982450574$  \\
\hline
$1.6 \times 10^{-6}$     & $2.6 \times 10^{-11}$  & $3.5 \times 10^{-7}$ & $3 \times 10^{-7}$  & $-0.499999464678626$ & $\pm 0.866024982450545$  \\
$1.6 \times 10^{-6}$     & $2.6 \times 10^{-11}$  & $3.5 \times 10^{-7}$ & $3 \times 10^{-5}$  & $-0.499999464663069$ & $\pm 0.866013755215885$  \\
$1.6 \times 10^{-6}$     & $2.6 \times 10^{-11}$  & $3.5 \times 10^{-7}$ & $3 \times 10^{-9}$  & $-0.499999464678781$ & $\pm 0.866025094722156$  \\
\hline
\end{tabular}}
\label{tab:eqt}
\end{center}
\end{table}

The position of non-collinear equilibrium points is calculated from Eq. \ref{eq:noncol_eq_fix}. Table \ref{tab:eqt} shows the position of non-collinear equilibrium points with respect to the chosen value of several parameters. When there are no perturbing factors, the triangular points have the same coordinate as in the classical case. The inclusion of perturbation parameters has resulted in a shift in the location of non-collinear equilibrium points. The increment of $A$ makes the position of these equilibrium points closer to the small primary. In contrast, if we reduce $q$ or increase $M_b$, the position of equilibrium points is shifted toward the bigger primary. The position is closer to the bigger primary in line with the increase of $l$. 

\begin{figure*}
\centering
    \subfloat[L1\label{fig:eigen_L1}]{%
    \includegraphics[width=0.40\textwidth]{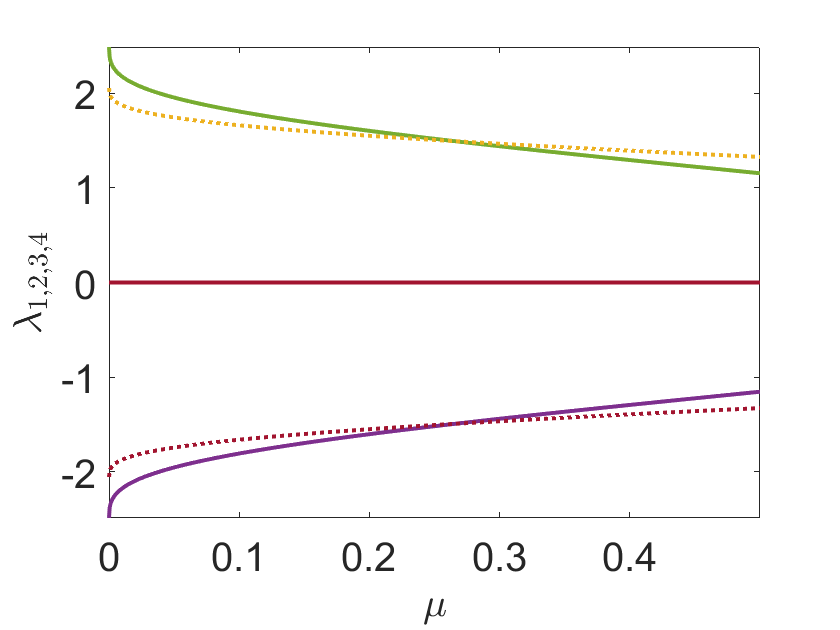}
    }
    \qquad
    \subfloat[L2\label{fig:eigen_L2}]{%
      \includegraphics[width=0.40\textwidth]{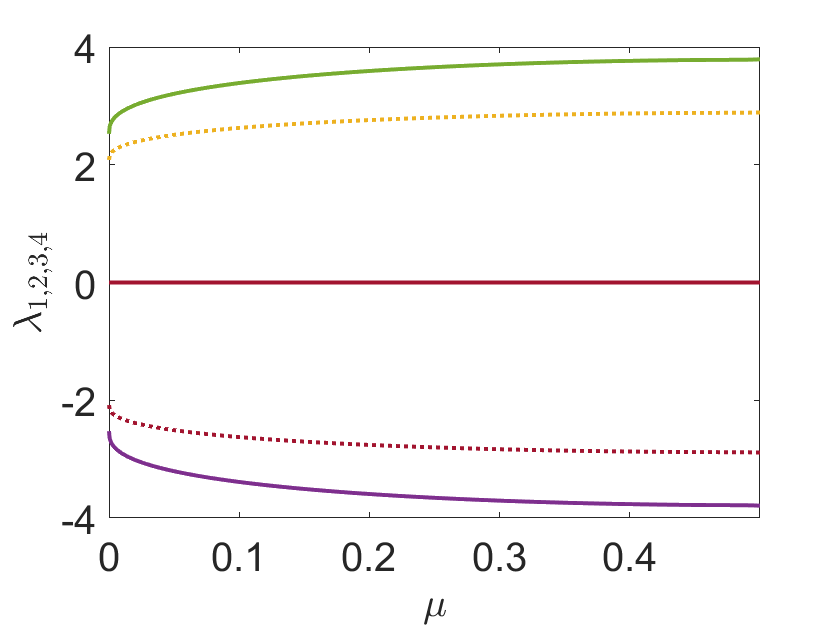}
    }
    \qquad
    \subfloat[L3\label{fig:eigen_L3}]{%
      \includegraphics[width=0.40\textwidth]{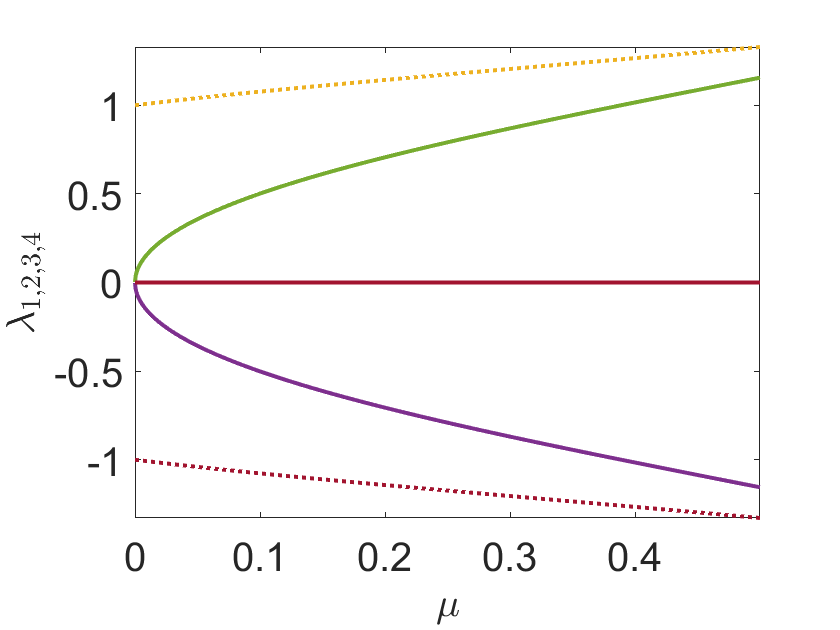}
    }
\caption{Plot of $\mu$ versus characteristic roots ($\lambda_{1,2,3,4}$) for L1, L2, and L3, with $l = 3.5 \times 10^{-7}$, $M_b = 3 \times 10^{-7}$, $A = 2.6 \times 10^{-11}$, and $ 1-q = 1.6 \times 10^{-6}$. The real and imaginary parts of characteristic roots are marked by solid and dashed lines, respectively. Here we used $T = 0.11$. }
\label{fig:eigen_col}
\end{figure*}

We now analyze the linear stability of each equilibrium point in the Sun-Haumea system. Collinear equilibrium points lie in the abscissa. Hence we have $\Omega^o_{xy} = 0$. In order to study the stability, we divide the abscissa into three regions, i.e., $L_1$ ($-\infty$, $\mu-1-l$), $L_2$ ($\mu-1-l$, $\mu$), and $L_3$ ($\mu$, $\infty$), and calculate the sign of $b$ and $b^2-4c$ numerically for each region. First, we estimate the stability by considering the perturbation parameters in the Sun-Haumea system. As shown in Figure \ref{fig:eigen_col}, there exist pure real and pure imaginary characteristic roots for $\mu$ between 0 and 0.5. Hence, all collinear equilibrium points of the Sun-Haumea system are unstable. Furthermore, we conducted the calculation by varying the value of perturbation parameters. Table \ref{tab:p_valeigen}  shows the result of the calculation. All regions have $b < 0$ and $b^2-4c > 0$ which means it produces two real pairs and two pure imaginary pairs. It shows that even if we change the value of perturbation parameters, the collinear equilibrium points remain unstable.

\begin{landscape}
\begin{table}
\caption{Characteristic roots of collinear equilibrium points in Sun-Haumea system with $\mu = 2 \times 10^{-9}$. Here $i$ means $\sqrt{-1}$. We used $T = 0.11$. }
\begin{center}
{\renewcommand{\arraystretch}{1.5}
\setlength{\tabcolsep}{4pt} 
\begin{tabular}{|c|c|c|c|c|c|c|c|c|c|}
\hline
\multirow{2}{*}{$1-q$}   & \multirow{2}{*}{$A$}   & \multirow{2}{*}{$l$}   & \multirow{2}{*}{$M_b$} &\multicolumn{2}{|c|}{$L_1$}&\multicolumn{2}{|c|}{$L_2$}&\multicolumn{2}{|c|}{$L_3$} \\ \cline{5-6} \cline{7-8} \cline{9-10} 
&&& & $\lambda_{1,2}$ & $\lambda_{3,4}$ & $\lambda_{1,2}$ & $\lambda_{3,4}$ & $\lambda_{1,2}$ & $\lambda_{3,4}$\\
\hline
1                        & 0                      & 0                    & 0                   & $2.50618628025287$ & $2.07031520790267i$ & $2.51039039369088$ & $2.07287538016581i$ & $0.000072456881366$ & $1.00000000175000i$  \\
\hline
$1.6 \times 10^{-6}$     & $2.6 \times 10^{-11}$  & $3.5 \times 10^{-7}$ & $3 \times 10^{-7}$  & $2.50732613894453$ & $2.07100940502565i$ & $2.50924970700464$ & $2.07218079713681i$ & $0.000074119040096$ & $1.00000030173167i$  \\
$1.6 \times 10^{-4}$     & $2.6 \times 10^{-11}$  & $3.5 \times 10^{-7}$ & $3 \times 10^{-7}$  & $2.58032763652133$ & $2.11561449848464i$ & $2.43683616339914$ & $2.02823606045270i$ & $0.000074121110154$ & $1.00000030173314i$  \\
$1.6 \times 10^{-9}$     & $2.6 \times 10^{-11}$  & $3.5 \times 10^{-7}$ & $3 \times 10^{-7}$  & $2.50659334481697$ & $2.07056321377481i$ & $2.50998449730882$ & $2.07262831852643i$ & $0.000074119018377$ & $1.00000030173165i$  \\
\hline
$1.6 \times 10^{-6}$     & $2.6 \times 10^{-11}$  & $3.5 \times 10^{-7}$ & $3 \times 10^{-7}$  & $2.50732613894453$ & $2.07100940502565i$ & $2.50924970700464$ & $2.07218079713681i$ & $0.000074119040096$ & $1.00000030173167i$  \\
$1.6 \times 10^{-6}$     & $2.6 \times 10^{-9}$   & $3.5 \times 10^{-7}$ & $3 \times 10^{-7}$  & $2.50732614822158$ & $2.07100941067903i$ & $2.50924971630782$ & $2.07218080279818i$ & $0.000074119037849$ & $1.00000029980116i$  \\
$1.6 \times 10^{-6}$     & $2.6 \times 10^{-13}$  & $3.5 \times 10^{-7}$ & $3 \times 10^{-7}$  & $2.50732613885112$ & $2.07100940496873i$ & $2.50924970691157$ & $2.07218079708017i$ & $0.000074119039347$ & $1.00000030175097i$  \\
\hline
$1.6 \times 10^{-6}$     & $2.6 \times 10^{-11}$  & $3.5 \times 10^{-7}$ & $3 \times 10^{-7}$  & $2.50732613894453$ & $2.07100940502565i$ & $2.50924970700464$ & $2.07218079713681i$ & $0.000074119040096$ & $1.00000030173167i$  \\
$1.6 \times 10^{-6}$     & $2.6 \times 10^{-11}$  & $3.5 \times 10^{-5}$ & $3 \times 10^{-7}$  & $2.50925730691060$ & $2.07218542599087i$ & $2.51118123266234$ & $2.07335725400133i$ & $0.000074119034105$ & $1.00000030234410i$  \\
$1.6 \times 10^{-6}$     & $2.6 \times 10^{-11}$  & $3.5 \times 10^{-9}$ & $3 \times 10^{-7}$  & $2.50732594585502$ & $2.07100928745113i$ & $2.50924951387951$ & $2.07218067951878i$ & $0.000074119040096$ & $1.00000030173161i$  \\
\hline
$1.6 \times 10^{-6}$     & $2.6 \times 10^{-11}$  & $3.5 \times 10^{-7}$ & $3 \times 10^{-7}$  & $2.50732613894453$ & $2.07100940502565i$ & $2.50924970700464$ & $2.07218079713681i$ & $0.000074119040096$ & $1.00000030173167i$  \\
$1.6 \times 10^{-6}$     & $2.6 \times 10^{-11}$  & $3.5 \times 10^{-7}$ & $3 \times 10^{-5}$  & $2.54765547950673$ & $2.09562955258651i$ & $2.46924636836900$ & $2.04788142310119i$ & $0.000172088738229$ & $1.00003000142140i$  \\
$1.6 \times 10^{-6}$     & $2.6 \times 10^{-11}$  & $3.5 \times 10^{-7}$ & $3 \times 10^{-9}$  & $2.50692401885402$ & $2.07076439389322i$ & $2.50965096408882$ & $2.07242501730710i$ & $0.000072473711215$ & $1.00000000473057i$  \\
\hline

\hline
\end{tabular}}

\label{tab:p_valeigen}
\end{center}
\end{table}
\end{landscape}

\begin{figure*}
\centering
    \subfloat[\label{fig:eigen_00}]{%
    \includegraphics[width=0.43\textwidth]{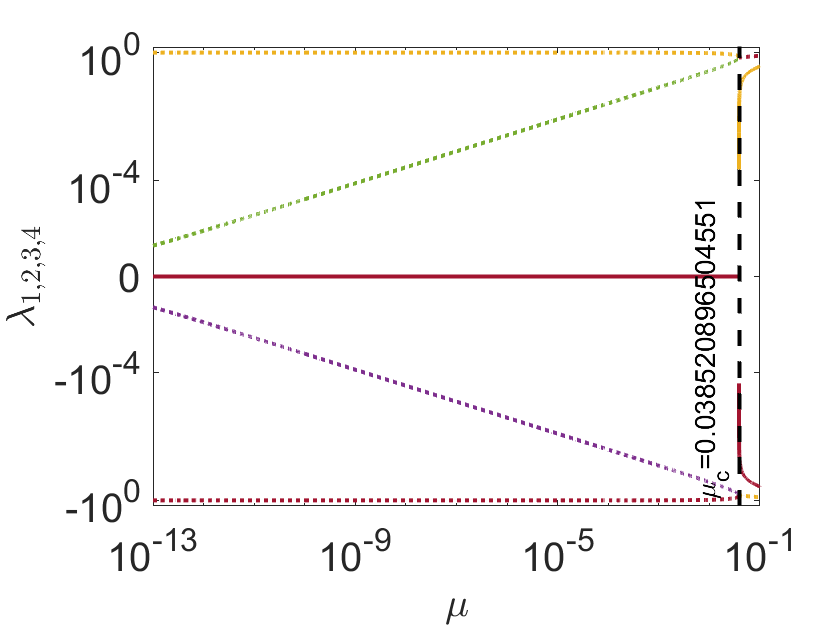}
    }
    \qquad
    \subfloat[ \label{fig:eigen_0}]{%
      \includegraphics[width=0.43\textwidth]{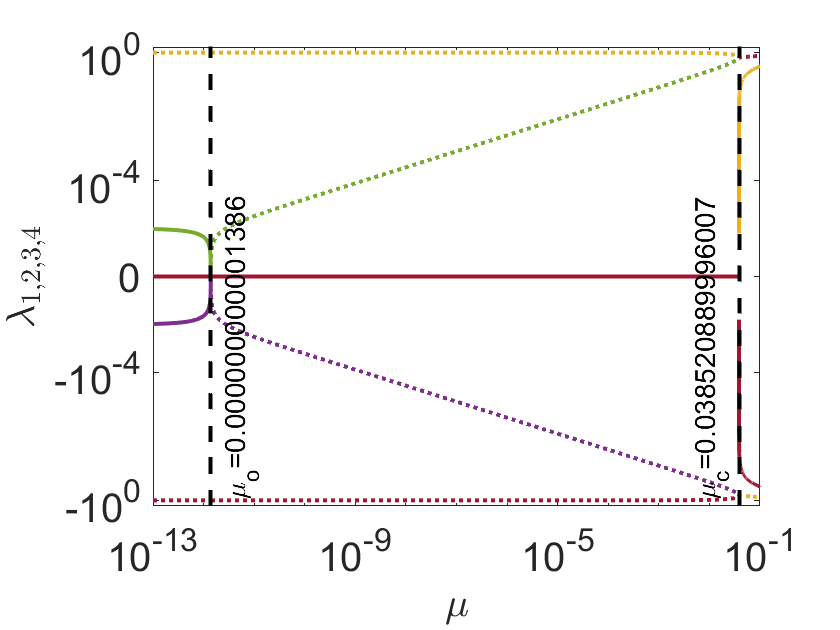}
    }
    \hfill
    \subfloat[\label{fig:eigen_A}]{%
      \includegraphics[width=0.43\textwidth]{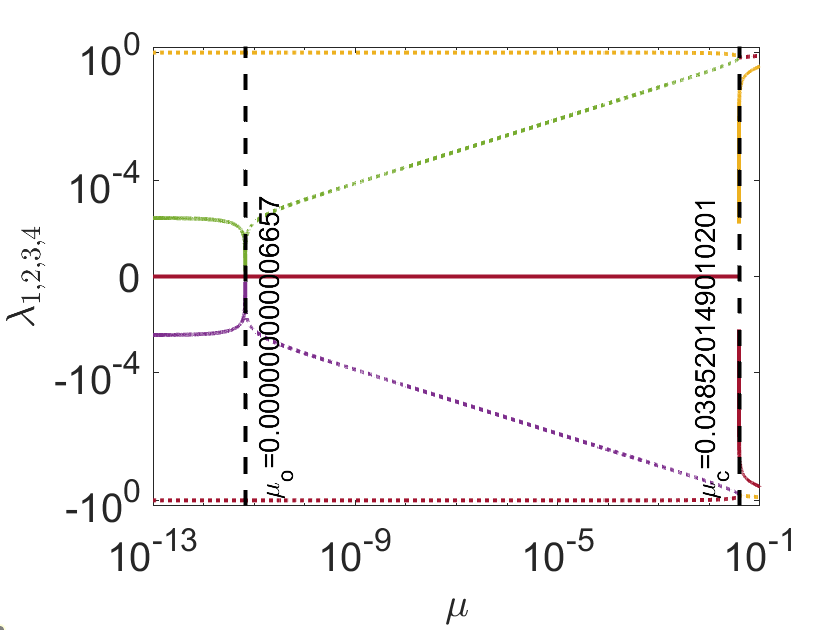}
    }
    \qquad
    \subfloat[\label{fig:eigen_q}]{%
      \includegraphics[width=0.43\textwidth]{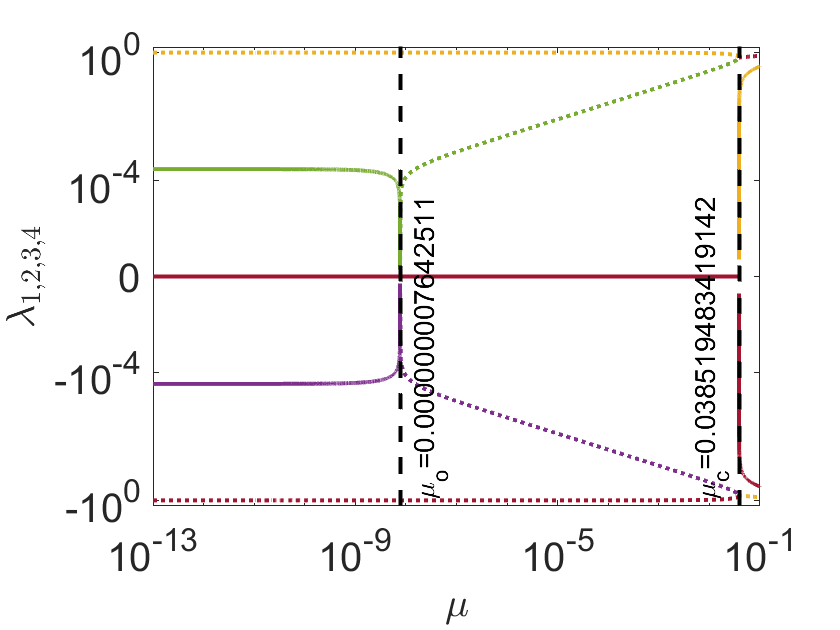}
    }
    \hfill
    \subfloat[\label{fig:eigen_Mb}]{%
      \includegraphics[width=0.43\textwidth]{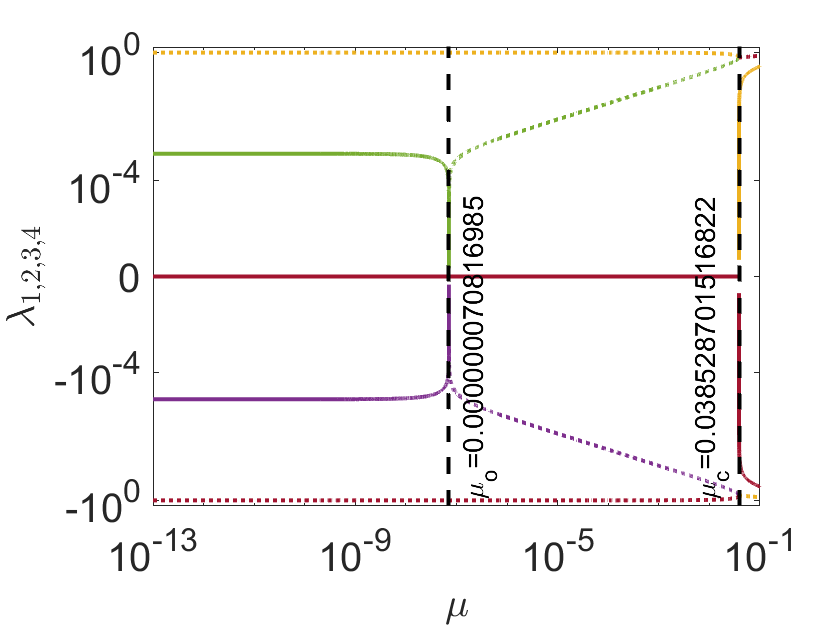}
    }
    \qquad
    \subfloat[\label{fig:eigen_l}]{%
      \includegraphics[width=0.43\textwidth]{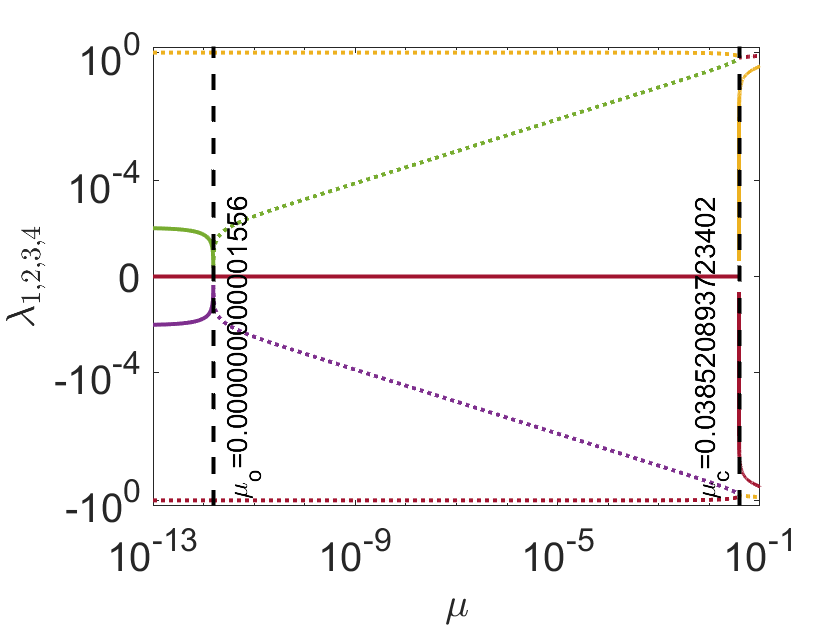}
    }
\caption{Plot of $\mu$ versus characteristic roots ($\lambda_{1,2,3,4}$) in L4 with different parameters configuration. The real and imaginary parts of characteristic roots are marked by solid and dashed lines respectively. Here we used $T = 0.11$. The detail of the parameters that are used in each subfigure is as follows. (a) $A = 0$, $1-q = 0$, $l = 0$, $M_b = 0$. (b) $A = 2.6 \times 10^{-11}$, $1-q = 1.6 \times 10^{-6}$, $l = 3.5\times 10^{-7}$, $M_b = 3\times 10^{-7}$. (c) $A = 2.6 \times 10^{-6}$, $1-q = 1.6 \times 10^{-6}$, $l = 3.5\times 10^{-7}$, $ M_b = 3\times 10^{-7}$. (d) $A = 2.6 \times 10^{-11}$, $1-q = 1.6 \times 10^{-4}$, $l = 3.5\times 10^{-7}$, $M_b = 3\times 10^{-7}$. (e) $A = 2.6 \times 10^{-11}$, $1-q = 1.6 \times 10^{-6}$, $l = 3.5\times 10^{-7}$, $M_b = 3\times 10^{-4}$. (f) $A = 2.6 \times 10^{-11}$, $1-q = 1.6 \times 10^{-6}$, $l = 3.5\times 10^{-4}$, $M_b = 3\times 10^{-7}$.}
\label{fig:stability}
\end{figure*}

Next, we investigate the stability of non-collinear equilibrium points in the Sun-Haumea system. We discuss only $L_4$ since the dynamic of $L_5$ is nearly similar. In the classical case, non-collinear equilibrium points are stable under the condition $27\mu(1-\mu) < 1$. Hence we can deduce $\mu < \mu_c$, where the critical mass $\mu_c = 0.038520896504551$. This critical mass can be calculated by finding the solution of $b^2 - 4c = 0$. In this modified version of CRTBP, we numerically calculate the roots by solving Eq. \ref{eq:characteristic}. By considering the perturbing parameters, it shows that the stability of non-collinear equilibrium points has a maximum limit ($\mu_c$) and minimum limit ($\mu_o$) of mass parameters which is different from the classical case. For Sun-Haumea system, we found $\mu_c = 0.0385208896007$ and $\mu_o = 1.386 \times 10^{-12}$. Since the Sun-Haumea system has $\mu = 2 \times 10^{-9}$, we conclude that the Sun-Haumea system has stable non-collinear equilibrium points. Figure \ref{fig:stability} shows the comparison of stability for several cases by changing the perturbing parameters of the Sun-Haumea system. It shows that the range of stability depends on the parameter $A$, $q$, $l$, and $M_b$.  The characteristic roots have the form of pure imaginary if $\mu_o < \mu < \mu_c$. The considered perturbation parameters alter the range of stability in $\mu$. The increment of $A$ or reduction of $q$ reduces the size of the stability area. The stability region is shifted toward bigger $\mu$ if $M_b$ and $l$ increase.

\section{Conclusion}\label{sec5}

We have investigated the dynamics of an infinitesimal mass under the gravitational influence of two primaries. Our study assumes that the smaller primary is an elongated body, while the larger primary is oblate and also emits radiation. In addition, we have taken into account the presence of a disk that surrounds the three-body system. We have found that there are five equilibrium points in this modified CRTBP where three of them are collinear and the other two are non-collinear. Our numerical exploration of the Sun-Haumea system has revealed that the inclusion of perturbing parameters has caused a displacement in the position of the Sun-Haumea system's equilibrium points with respect to their positions in the classical CRTBP. We noticed that the magnitude of the perturbing parameters ($q$, $A$, $l$, and $M_b$) can affect the position of the five equilibrium points. It shows that the non-collinear equilibrium points of the Sun-Haumea system are stable, while all collinear equilibrium points are unstable. Moreover, we have figured out that the collinear equilibrium points remain unstable for several possible ranges of perturbing parameters. In contrast, the non-collinear equilibrium points are conditionally stable with respect to $\mu$. When taking into account the perturbing parameters, we have found that there are upper and lower limits of $\mu$ for achieving the stability of non-collinear equilibrium points. The stability region in $\mu$ depends on the perturbing parameters. 

\normalem
\begin{acknowledgements}
This work is funded partially by BRIN's research grant Rumah Program AIBDTK 2023. We thank the anonymous reviewer for the insightful comments and suggestions on the manuscript. 
\end{acknowledgements}

\bibliographystyle{raa}
\bibliography{nur}

\end{document}